# Anomalous enhancement of nanodiamond luminescence on heating

A A Khomich[1,2], O S Kudryavtsev[1], T A Dolenko[3,4], A A Shiryaev[5,6], A V Fisenko[7], V I Konov[1,4] and I I Vlasov[1,4]

[1] Prokhorov General Physics Institute RAS, 119991, Moscow, Russia
[2] Kotelnikov Institute of Radio Engineering and Electronics RAS, 141190, Fryazino, Moscow region, Russia
[3] Lomonosov Moscow State University, 119991, Moscow, Russia
[4] National Research Nuclear University 'MEPhI', 115409, Moscow, Russia
[5] Institute of physical chemistry and electrochemistry RAS, 119071, Moscow, Russia
[6] Institute of Geology of Ore Deposits, Petrography, Geochemistry and Mineralogy RAS, 119071, Moscow, Russia
[7] Vernadsky Institute of Geochemistry and Analytical Chemistry RAS, 119991, Moscow, Russia

E-mail: vlasov@nsc.gpi.ru

**Abstract**
Characteristic photoluminescence (PL) of nanodiamonds (ND) of different origin (detonation, HPHT, extracted from meteorite) was studied *in* situ at high temperatures in the range 20–450 °C. Luminescence was excited using 473 nm laser and recorded in the range 500–800 nm. In contrast to decrease of point defect PL in bulk diamond with temperature, we found that the nanodiamond luminescence related to ND surface defects increases almost an order of magnitude upon heating to 200–250 °C. The observed effect reveals that water adsorbed on ND surfaces efficiently quenches photoluminescence; water desorption on heating leads to dramatic increase of the radiative de-excitation.

## 1. Introduction

Broad-band photoluminescence (PL) in the range 400–700 nm is typical for nanodiamonds of various origins. Since its intensity increases with decreasing nanodiamond size and, consequently, with surface/bulk ratio [1-5], it is usually ascribed to surface defects of nanodiamond grains, in particular, $sp^2$-carbon structures. It was shown [6] that both broad-band luminescence and $sp^2$-carbon Raman lines decrease strongly already after 40 minutes of oxidation of 5 nm detonation diamonds in air. That result corroborates the assignment of the luminescence to nanodiamond shell containing $sp^2$-carbon.

Nanodiamond luminescence depends on functional state of the ND surface. For example, the maximum of the PL band can be moved toward UV range by covalent linking of octadecylamine to ND particles [7]. Intensity of the nanodiamond PL in aqueous suspension can be adjusted by employing functional groups having different strength of interaction with water [8].

In the present work a new property of the broad-band luminescence of dry nanodiamond powders is reported: considerable increase of the intensity upon air heating to 200–250 °C.

## 2. Experimental

Several types of nanodiamonds were studied: a) detonation nanodiamonds (DND), b) mechanically grinded synthetic High Pressure – High Temperature nanodiamonds (HPHT ND), c) nanodiamonds extracted from meteorites (MND) (see Table 1). DND were produced by detonation of trinitrotoluene and 1,3,5- trinitrotoluene-1,3,5-triazine mixture in medium with gas cooling (SIA Altay, Russia) and subsequently cleaned in a mixture of sulfuric and nitric acids

(Adamas Nanotechnologies, USA). HPHT ND were provided by Tomei Diamond, Japan. Meteoritic ND were extracted from Krymka meteorite according to protocol described in [9, and ref. therein].

Photoluminescence spectra of the nanodiamond powders were recorded using LABRAM HR800 spectrometer equipped with the diode-pumped solid-state laser Ciel-473 (Laser Quantum). The laser beam (power 0.1-1.0 mW; wavelength 473 nm) was focused in a 2 μm spot on ND powder surface.

Photoluminescence spectra were recorded *in situ* in the Linkam TS 1500 heating stage in the temperature range 20–450 °C. Temperature step was 25 °C; dwell time at a given temperature – 25 min. The samples were prepared by drying of ND aqueous suspension on Si substrate till formation of continuous film. All spectra were recorded in 477–800 nm, range; duration of each acquisition was approx. 5 min.

Composition of surface-bound functional groups was addressed with Infra-Red (IR) microscope Auto Image connected to Spectrum One spectrometer (Perkin Elmer); high temperature measurements were also performed with the Linkam stage. The samples were prepared by drying of ND suspension on Al mirror. Obtained spectra represent superposition of transmission and reflectance spectra with strong domination of the former. This approach permits to minimize eventual hydrogenation of nanodiamond grains during sample preparation.

**Table 1.** List of the tested nanodiamonds indicating their characteristic sizes and concentrations of main impurities (nitrogen and silicon).

| Type of nanodiamonds (ND) | Grain size (nm) | Concentration of impurity in the bulk |
|---|---|---|
| Detonation ND | 5 | 10 000 ppm N |
| Grinded HPHT ND | 10-20 | 100 ppm N |
| Meteoritic ND | 1-10 | 10 000 ppm N + ? ppm Si |

### 3. Results and discussion

At room temperature all tested ND samples possess qualitatively similar luminescence spectra with featureless broad band in 500–700 nm (Figure 1).

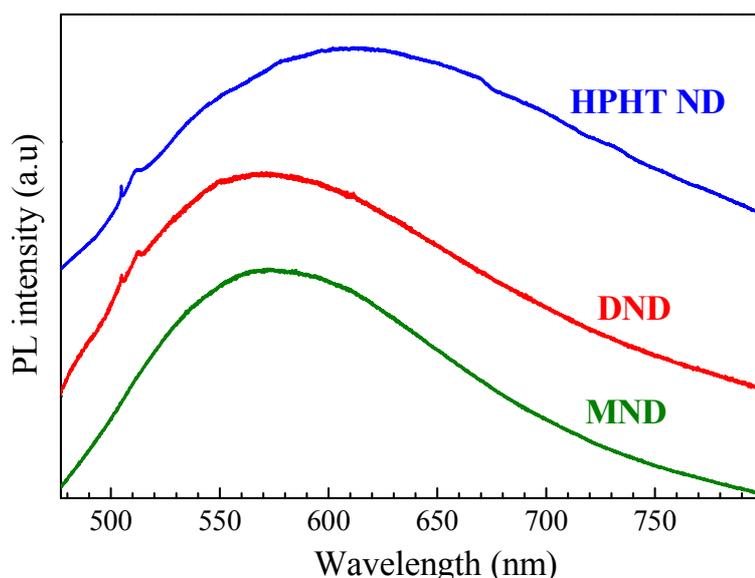

**Figure 1.** Room temperature PL spectra of the tested nanodiamonds; excitation at 473 nm.

In the following we discuss behavior of integral PL intensity, calculated as an area under the spectrum in the range 470–800 nm. Figure 2 shows temperature dependencies of integrated PL for DND, HPHT and MND samples. For all samples an unusual behavior is observed: after expected slight initial decrease at 50–75 °C, the integrated PL intensity increases with temperature, peaks at ~200–250 °C and then decreases. After cooling tested sample down to 20 °C and starting a new heating cycle the observed PL temperature dependence is completely repeated.

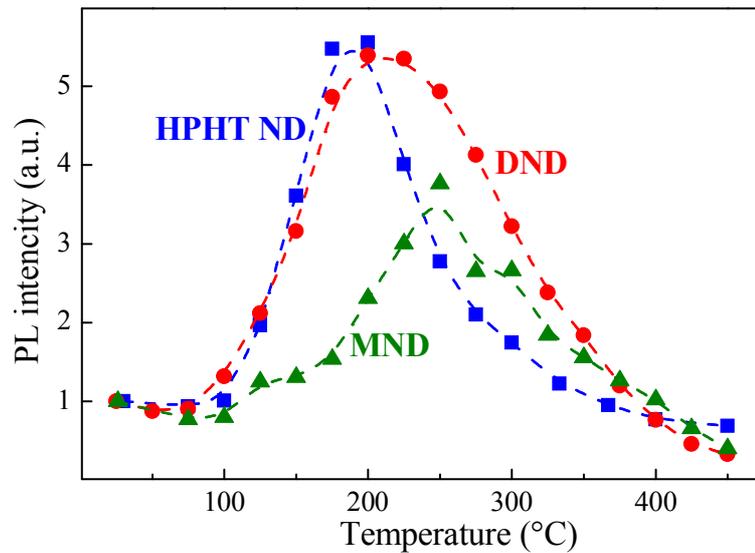

**Figure 2.** The dependences of integrated (in the range 470–800 nm) photoluminescence on heating temperature for the HPHT ND (square points), DND (circle points), MND (triangle points) samples.

Some sample-dependent differences in the temperature dependence of PL intensity are observed. In comparison with DND and MND samples, the HPHT ND is characterized by somewhat lower temperature and higher intensity of the maximum and by steeper decrease of the intensity at high temperatures. The MND sample is characterized by the highest T of the maximum and its intensity is the weakest. Such a difference could be explained by specific morphology of each sample. HPHT ND consists of individual crystallites, whereas an aggregated form of diamond crystallites is more typical for DND and MND. The ND aggregation slows the effects of temperature on functional state of the crystallite surface, and, as a consequence, on the change of the PL intensity.

The observed temperature dependence of photoluminescence drastically differs from behavior of bulk PL of various materials and, in particular, from PL related to point defects in bulk diamond. Photoluminescence intensity usually decreases with temperature due to increased probability of non-radiative relaxation of excited electrons, appearance of electron traps, etc. Figure 3 shows typical temperature dependence of luminescence of "nitrogen-vacancy" (NV) defects in bulk diamond.

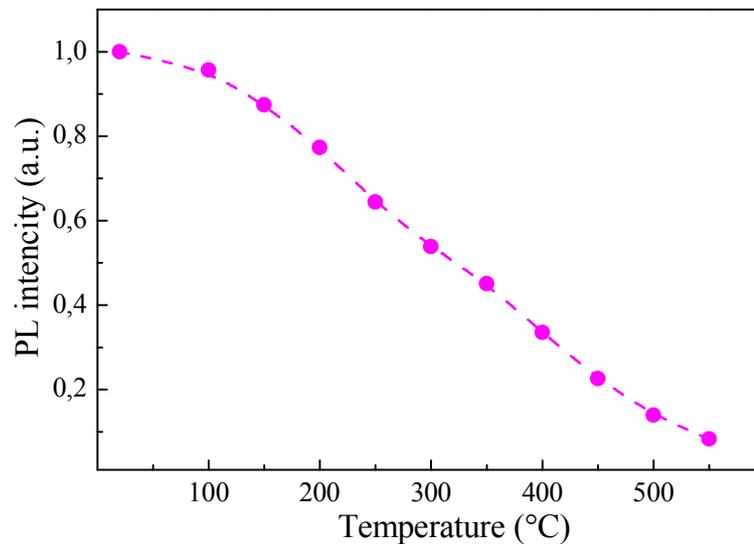

**Figure 3.** Temperature quenching of photoluminescence of NV-defects in bulk diamond.

Qualitatively similar temperature behaviour of integral luminescence (thermoluminescence) was earlier observed for meteoritic and synthetic nanodiamonds after priming with X-rays [10, 11]. However, the effect observed in this work cannot be attributed to thermoluminescence, since in our case no priming was employed and, more important, in contrast to thermoluminescence, the temperature dependence of the photoluminescence is multiply repeated after cooling of NDs in air.

It is known that water, adsorbed on surfaces of nanodiamond grains and in pores is removed by heating to 150–200 °C [12], though surface OH-groups could survive to much higher temperatures [13]. This allows to propose that the observed unusual behavior of photoluminescence on heating could be related to fate of adsorbed water. We suggest that water efficiently quenches surface-related luminescence of nanodiamonds at room temperature; its removal during the heating intensifies the PL and at yet higher temperatures conventional quenching mechanism becomes dominant. To verify this hypothesis, we have studied temperature dependence of the PL from dehydrated nanodiamonds by performing heating experiments in dry nitrogen. In this case the PL intensity at room temperature was approximately 5 times stronger than that of the starting powders and steadily decreased with temperature, thus following "usual" behavior.

Additional support to the model comes from Infra-red spectroscopic study of functional groups of ND surfaces at different temperatures. It is shown (Figure 4) that intensity of absorption bands at 3000–3600 and 1630 $cm^{-1}$, corresponding to valence and deformation vibration modes of water molecule, respectively, notably decreases at ~200 °C and remains virtually constant at higher temperatures. The remaining absorption bands in the range 3000–3600 $cm^{-1}$ is likely related to vibrations of surface OH-groups. Note that intensity of absorption bands unrelated to HOH and OH vibrations is unaffected by the heating. Spectra of ND samples cooled in air after the heating show gradual recovery of the water-related bands due to adsorption of atmospheric moisture.

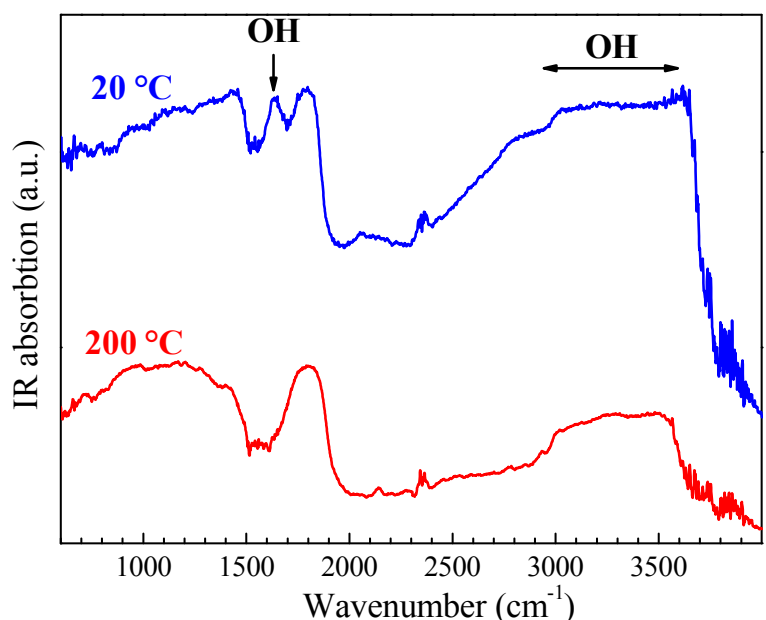

**Figure 4.** IR absorption spectra of HPHT ND at 20 °C and 200 °C temperatures. The curves are shifted vertically for clarity. Horizontal and vertical arrows show, respectively, positions of valence and deformation vibration modes of water molecules.

Last two experiments demonstrated close correlation of integral PL intensity of nanodiamonds with water desorption process. Absence of the PL enhancement on heating of dehydrogenated nanodiamonds and notable decrease of water bands in IR absorption spectra of ND, heated up to 200 °C, confirm that interaction of water with surface luminescing defects is responsible for the observed unusual temperature dependence of the integral PL.

Water is one of the most important adsorbates on various carbons [14], In particular, nanodiamonds effectively adsorb OH-groups and molecular water both by physical and chemical mechanisms [e.g., 15]. However, correlation between the amounts of adsorbed water and luminescence properties was not previously reported. In our view, the most plausible explanation of the PL quenching by adsorbed water is related to proton or electron transport between the luminophore and water. The probability of such charge transport increases in the presence of strong hydrogen bonds between ND surface groups and water, which was recently found for nanodiamonds [16]. The charge transfer follows a laser excitation of luminophore and often leads to luminescence quenching [17].

**4. Conclusions**

The broad-band photoluminescence of nanodiamonds of different types was investigated under their heating *in situ* in the temperature range 20–450 °C. It is shown that intensity of the photoluminescence is the highest at 200–250 °C (exact value is sample-dependent) and is approx. 5 times stronger than at room temperature (20 °C). Dehydrogenated nanodiamonds do not possess such behavior. These observations are explained by desorption of water from grain surfaces upon heating.

Deep understanding of luminescence properties of nanodiamonds is necessary for possibility to control them, in particular, to change the intensity and/or shift the band position. Applications of nanodiamonds such as luminescing biomarkers [18], single-photon emitters [19], high-sensitivity magnetometers [20, 21], and temperature sensors [22] are among fields, which would greatly benefit from discovery of means to control the yield of the surface-related background luminescence.

**Acknowledgments**

The research was supported by grant of Russian Science Foundation (grant No 14-12-01329) in studies of synthetic nanodiamonds and by RFBR grant No 15-05-03351 in studies of meteoritic nanodiamonds.